\documentclass[iop]{emulateapj}
\shorttitle{The Aquarius Dwarf Galaxy}
\shortauthors{Cole et al.}
\begin{document}
\title{Delayed Star Formation in Isolated Dwarf Galaxies: HST Star Formation History 
of the Aquarius Dwarf Irregular\altaffilmark{1}}
\author{Andrew A. Cole\altaffilmark{2}, Daniel R. Weisz\altaffilmark{3,4,11},
Andrew E. Dolphin\altaffilmark{5}, Evan D. Skillman\altaffilmark{6},
Alan W. McConnachie\altaffilmark{7}, Alyson M. Brooks\altaffilmark{8},
Ryan Leaman\altaffilmark{9,10} }
\altaffiltext{1}{Based on observations made with the NASA/ESA Hubble Space Telesope,
obtained at the Space Telescope Science Institute, which is operated by the Association
of Universities for Research in Astronomy, Inc., under NASA contract NAS 5-26555. These
observations were obtained under program GO-12925.}
\altaffiltext{2}{School of Physical Sciences, University of Tasmania, Private Bag 37, Hobart, 
Tasmania, 7001 Australia; andrew.cole@utas.edu.au}
\altaffiltext{3}{Department of Astronomy, University of California at Santa Cruz, 1156 High Street,
Santa Cruz, CA 95064; drw@ucsc.edu}
\altaffiltext{4}{Department of Astronomy, University of Washington, Box 351580, Seattle, WA 98195 USA}
\altaffiltext{5}{Raytheon; 1151 E. Hermans Rd, Tucson, AZ 85706, USA; adolphin@raytheon.com}
\altaffiltext{6}{Minnesota Institute for Astrophysics, University of Minnesota, Minneapolis, MN 55441, USA; skillman@astro.umn.edu}
\altaffiltext{7}{NRC Herzberg Institute of Astrophysics, Dominion Astrophysical Observatory, Victoria, BC, V9E~2E7 Canada; alan.mcconnachie@nrc-cnrc.gc.ca}
\altaffiltext{8}{Rutgers, the State University of New Jersey, Department of Physics \& Astronomy, 136 Frelinghuysen Rd, Piscataway, NJ 08854, USA; abrooks@physics.rutgers.edu}
\altaffiltext{9}{Instituto de Astrof\'{\i}sica de Canarias, E-38205 La~Laguna, Tenerife, Spain; rleaman@iac.es}
\altaffiltext{10}{Departamento de Astrof\'{\i}sica, Universidad de La~Laguna, E-38206 La~Laguna, Tenerife, Spain}
\altaffiltext{11}{Hubble Fellow}
\begin{abstract}
We have obtained deep images of the highly isolated (d = 1~Mpc) Aquarius dwarf irregular galaxy (DDO~210) with the Hubble Space Telescope (HST) Advanced Camera for Surveys (ACS). The resulting color-magnitude diagram (CMD) reaches more than a magnitude below the oldest main-sequence turnoff, allowing us to derive the star formation history (SFH) over the entire lifetime of the galaxy with a timing precision of $\approx$10\% of the lookback time. Using a maximum likelihood fit to the CMD we find that only $\approx$10\% of all star formation in Aquarius took place more than 10~Gyr ago (lookback time equivalent to redshift $z \approx2$). The star formation rate increased dramatically $\approx$6--8~Gyr ago ($z \approx$0.7--1.1) and then declined  until the present time.  The only known galaxy with a more extreme confirmed delay in star formation is Leo~A, a galaxy of similar M$_{HI}$/M$_{\star}$, dynamical mass, mean metallicity, and degree of isolation. The delayed stellar mass growth in these galaxies does not track the mean dark matter accretion rate from CDM simulations.  The similarities between Leo~A and Aquarius suggest that if gas is not removed from dwarf galaxies by interactions or feedback, it can linger for several gigayears without cooling in sufficient quantity to form stars efficiently. We discuss possible causes for the delay in star formation including suppression by reionization and late-time mergers. We find reasonable agreement between our measured SFHs and select cosmological simulations of isolated dwarfs. Because star formation and merger processes are both stochastic in nature, delayed star formation in various degees is predicted to be a characteristic (but not a universal) feature of isolated small galaxies.
\end{abstract}
\section{Introduction}

\subsection{Environment and the Evolution of Dwarf Galaxies}

The question of the early evolution of low-mass galaxies has been a vexing one for galaxy formation theory in a wide range of hierarchical merger models for more than three decades \citep[e.g.,][]{dek86,kau93,som02,ben03,saw11,kuh12}. During this timespan a range of problems has been identified and addressed in successive generations of models; these problems tend to relate to the efficiency with which gas is accreted, retained, cooled, and astrated in the lowest-mass dark matter halos. 

Recurring issues surrounding the over- or under-production of stars in simulations include overcooling, reionization, the role of feedback, and the effects of tidal perturbations; a recent, detailed discussion of these issues in the context of Local Group dwarf galaxies may be found in \citet{ski14}. A variety of recently proposed solutions invoke a range of external heating and internal feedback processes, many of which are strongly mass and/or environment dependent \citep[e.g.,][and references therein]{hop13}; the influence of numerical resolution on the simulations is also likely to play a role \citep[e.g.,][]{bro14}. In addition to supernova feedback, reionization of the Universe at redshifts $\gtrsim$6 has been considered as a source of  heating that might inhibit star formation in the smallest galaxies \citep[e.g.,][]{bab92}.  In the simplest formulation, reionization may quench star formation in galaxies with total dark halo mass $\lesssim$10$^{10}$ M$_{\odot}$ by warming gas to the point where it may be driven out of the shallow potential wells of these systems \citep[e.g.,][]{tho96,gne00,ric02}; a review of the observational evidence for and against such a scenario has been given by \citet{wei14b}, who find that there is some evidence for negative feedback on star formation, but with large galaxy-to-galaxy variability and with some inconsistency between the quenching timescale and the epoch of reionization.

While some process (or set of processes) clearly inhibits the formation of dwarf galaxies compared to their proliferation in  cold dark matter-only simulations, the nature of these processes is still an open question. Continued observational study of {\it isolated} dwarfs is necessary for at least two reasons; one is to probe the ancient star formation history independent of local external perturbations, and another is to investigate the possible quenching or delay in star formation near the transition mass below which star formation is strongly suppressed.  The most isolated galaxy previously observed with sufficient depth and resolution to distinguish the oldest main-sequence turnoff (MSTO) population is Leo~A \citep[DDO~69;][]{col07}, a galaxy on the opposite side of the Local Group from Aquarius, but with similar dynamical mass. Leo~A was found to be the dwarf galaxy with the youngest mean age and the smallest fraction of stars older than 10~Gyr of any Local Group member \citep[][and references therein]{hid13,wei14a}.

In this context, it is of great interest to know if the unusual SFH of Leo~A presented in \cite{col07} is an anomaly or a general feature of small, isolated galaxies. In this paper we show that the Aquarius dwarf has many similarities with Leo~A, and explore some of the implications of this consistency. We present our observational program and data reductions in section~\ref{sec:obs}, and the stellar content of Aquarius is shown in \S\ref{sec:pops}. We derive the SFH of Aquarius in \S\ref{sec:sfh} and compare its initial low level of star formation and sudden increase some 6--8~Gyr ago to other Local Group dwarf galaxies in section~\ref{sec:disc}. We put these reults into context with recent galaxy formation models in \S\ref{sec:models}.

\subsection{The Aquarius Dwarf Irregular, DDO 210}

The Aquarius dwarf galaxy was discovered on Palomar sky survey prints by \citet{van59} and identified
by him as a low surface brightness, very blue galaxy at the limit of resolution into stars. 
Based on the radial velocity of neutral hydrogen, \citet{fis75} identified Aquarius as a possible Local Group member.
Aquarius is extremely isolated; \citet{mcc12} lists it as one of just three Local Group galaxies with a free-fall time to the Local Group barycenter larger than a Hubble Time given its radial velocity\footnote{The others being the Sagittarius dwarf irregular (SagDIG, ESO~594-4) and VV124 (UGC~4879)}. Currently only two other galaxies are known to lie within $\approx$500~kpc of Aquarius; these are SagDIG at a distance of $\approx$340~kpc, and NGC~6822, $\approx$440~kpc away. Numerical action models of Local Group dynamics indicate that Aquarius is on its first infall into the Local Group from quite a large distance, and is exceedingly unlikely to have tidally interacted with any other known galaxy \citep*[e.g.,][]{sha13}.

Early CCD photometric studies found Aquarius to have an extended history of star formation continuing to very recent times, albeit at a low level \citep{gre93}. Despite a high fraction of cold HI \citep{you03}, virtually no current star formation was detected by H$\alpha$ imaging surveys \citep*{van97}. The consequent lack of bright, evolved stars \citep[e.g.,][]{eli85,mar90} confirms Aquarius to be one of the lowest-luminosity gas-rich galaxies in the Local Group (and also resulted in longstanding ambiguity about its true distance).  Aquarius is classified as a transition type galaxy, with properties intermediate between the dwarf irregulars and spheroidals \citep[e.g.,][]{mat98}. \citet{van94} drew attention to a possible link between its low stellar mass, high gas fraction, and isolated location in the context of the Local Group density-morphology relation \citep{ein74}. The prospective connection to environment was further explored by \citet*{gre03} in the light of proposed gas-stripping mechanisms that might produce spheroidals from low-luminosity irregulars.

\begin{deluxetable}{lcc}
 \tablecolumns{3}
 \tablewidth{0pt} 
 \tablecaption{Basic Properties of Aquarius \label{tab:props}}
 \tablehead{
   \colhead{Parameter} & 
   \colhead{Value} &
   \colhead{References}}
 \startdata
  Galactic coordinates ($\ell, b$) (deg.) & 34$\fdg$1, $-$31$\fdg$4 & (1) \\  
  Distance   (kpc)                                   & 977 $\pm$45 & (2) \\
  Line of sight reddening E(B$-$V) (mag) & 0.045 & (3) \\
  Absolute magnitude M$_V$                   &  $-$10.6 & (4) \\
  Half-light radius (arcmin) & 1.47 $\pm$0.04 & (4) \\
  Mass within r$_{1/2}$ (M$_{\odot}$) & 1.8$^{+0.9}_{-0.7}\times10^{7}$ & (5,2) \\
  Stellar mass (M$_{\odot}$) & 1--2$\times$10$^6$ & (6,4,2) \\
  M$_{HI}$ (M$_{\odot}$) & 2.7$\times$10$^6$ & (7) \\
  current SFR (M$_{\sun}$ yr$^{-1}$) & 1.6$\times$10$^{-4}$ & (8) \\
  Stellar metallicity, Range (dex) & $-$1.44, $\pm$0.35 & (9) 
 \enddata
 \tablecomments{Key to references: (1) \citet{mar90}; (2) this paper; (3) \citet{sch11}; 
 (4) \citet{mcc06}; (5) \citet{kir14}; (6) \citet{lee99}; (7) \citet{beg06}; (8) \citet{lee09}; (9) \citet{kir13}.}
\end{deluxetable}

The basic properties of DDO~210 are summarized in Table~\ref{tab:props}. 
The first tip of the red giant branch (TRGB) distance to Aquarius, 950 $\pm$50~kpc, was published by \citet{lee99}, who confirmed both the Local Group membership and the unusually low luminosity of the galaxy. This distance was confirmed with HST/WFPC2 photometry by \citet{kar02}, who derived a distance of 940 $\pm$40~kpc; later reanalyses found distances of 
968$\pm$63 \citep{jac09} and 977$\pm$63 \citep{tul09}. \citet{mcc05} found two possible locations for the TRGB and favored the more distant value of 1071 $\pm$39~kpc, arguing that smaller estimates could be the result of contamination of 
the brightest part of the red giant branch (RGB) by asymptotic giant branch (AGB) stars. We adopted the more conservative distance in planning the present observations, but in the course of CMD modeling the best-fit solutions to the SFH were found to converge on a distance modulus of 24.95 $\pm$0.10; this corresponds to a distance of 977 $\pm$45~kpc and is in good agreement with our preliminary analysis of the TRGB magnitude in our data. A more complete reanalysis of the distance to Aquarius will be given in our analysis of the TRGB and variable star populations (Skillman et al., in preparation).

The first stellar metallicities were obtained by \citet{kir13}, who measured spectra in the region of the NIR calcium triplet for 24 red giants, finding a mean metallicity 3.6\% of solar (Z/Z$_{\sun}$ = 0.036, or expressed logarithmically and referred to the iron abundance, [Fe/H] =  $-$1.44).  This value is consistent with the expecations from the dwarf galaxy mass-metallicity relation of \citet{lee06} and typical for a dwarf galaxy of the luminosity of Aquarius. \citet{kir13} found a range of metallicities, with a 1$\sigma$ spread about the mean of $\pm$0.35 in the log, also typical for dwarfs with a prolonged history of star formation.

\begin{figure*}[t!]
\plotone{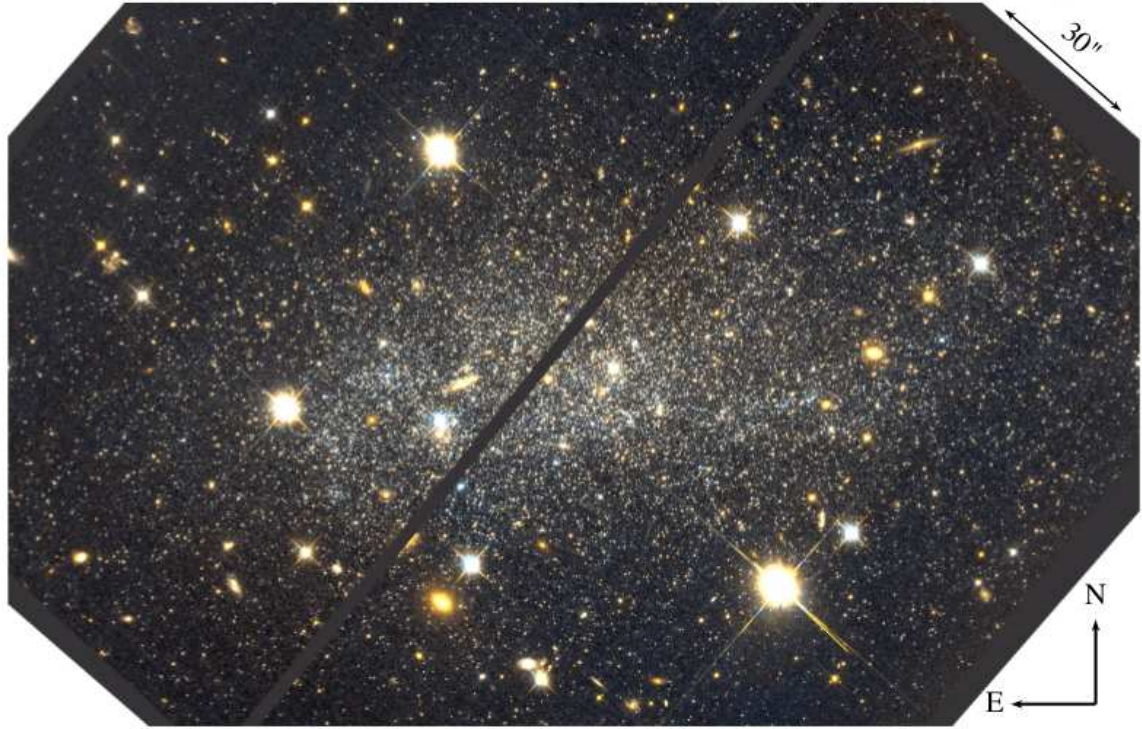}
\caption{The central 3$\farcm6\times2\farcm$3 portion of DDO210 as seen by ACS/WFC through the F475W (22,920 seconds) and F814W (33,480 seconds) filters. The 30$\arcsec$ scale bar corresponds to a physical length of 142~pc at the distance of DDO210.
\label{fig-image}}
\end{figure*}

\citet{kir14} measured the velocity dispersion of the Aquarius red giants to be $\sigma_v$ = 7.9$^{+1.9}_{-1.6}$~km~s$^{-1}$ and from this derived a dynamical mass at the half-light radius of 2.0$^{+1.0}_{-0.8}\times10^{7}$~M$_{\odot}$ and a mass to light ratio $\Upsilon$ = 32$^{+16}_{-14}$. This result is in good agreement with results based on the \ion{H}{1} rotation curve within 550~pc of the center of Aquarius, by \citet{beg06}. Under the assumption that the neutral hydrogen is distributed in a circularly symmetric disk with inclination $i_{HI} = 27^{\circ}$, \citet{beg06} found V$_{\mathrm{circ}} = 16$~km~s$^{-1}$, leading to an enclosed mass within 570~pc of M$_{\mathrm{dyn}} = 3.4\times10^7$ M$_{\odot}$. 

The two dynamical mass estimates agree well within the errors, recognizing that the half-light radius used by \citet{kir14} is 342~pc, about 62\% of the distance to the farthest point on the \ion{H}{1} rotation curve.  \citet{beg06} go farther, and match the velocity profile to a standard NFW dark matter profile \citep*{nfw}, finding a best fit maximum velocity at the virial radius V$_{200}$ $\approx$38~km~s$^{-1}$. This is a major extrapolation, given the virial radius of a few tens of kiloparsecs is many times farther out than the last measured data point. If taken at face value, this profile fit suggests a total mass within the virial radius on the order of 10$^{10}$~M$_{\odot}$, which allows at least rough comparisons to cosmological simulations.
 
\section{Observations \& Data Reduction} \label{sec:obs}
Aquarius was observed by HST under Cycle 20 program GO-12925 for 24 orbits spanning the time period between 2013 June 27 and June 30. Because of the low ecliptic latitude of Aquarius ($\beta$ = 4$\fdg$9) the observations were made when the angle between Aquarius and the Sun was near the optimum value of $\approx$135$\arcdeg$. The program was split into 12 visits of 2 orbits each in order to obtain enough separation between exposures to reliably identify short-period variable stars. We followed the Local Cosmology from Isolated Dwarfs (LCID) program \citep{mon10a} in using the F475W (Sloan $g^{\prime}$) and F814W (Broad~I) filters to optimally balance throughput and temperature sensitivity between the oldest MSTO stars and the base of the RGB. We used a subpixel dither pattern between each exposure in order to facilitate cosmic ray rejection, but did not attempt to fill in the gap between the two ACS/WFC chips since the fraction of stars lost in the gap is quite small.  The total exposure times in (F475W, F814W) were (22,920, 33,480)~seconds. 

A three-color composite image built from our co-added images is shown in Figure~\ref{fig-image}. This tends to highlight the brighter, bluer stars amongst the foreground stars and many background galaxies. No obvious nebulosities are present, consistent with the lack of current massive star formation in Aquarius. While the distribution of young stars is clumpy and irregular, the red giants are more smoothly distributed, and extend out beyond the boundaries of the ACS/WFC field \citep{mcc06}. We obtained parallel images with the WFC3 camera and did not convincingly detect any Aquarius members at that radius ($\gtrsim6\arcmin$), in agreement with the density profile derived by \citet{mcc06}.

The charge-transfer efficiency corrected images were processed through the standard HST pipeline and photometry was done using the most recent version of DOLPHOT \citep{dol00}. Extended objects and residual hot pixels were rejected based on their brightness profiles ($|$sharpness$|$ $<$0.32), and aperture corrections were derived based on relatively isolated stars picked from around the image. Stars that were found to suffer from excessive crowding noise (crowding parameter $>$1.0) due to partially resolved bright neighbors were rejected, leaving a sample of 51,239 well-measured stars with S/N $\geq$5.  We performed an artificial star test analysis with 500,000 stars in order to characterize the measurement errors and incompleteness. Artificial stars with the appropriate point spread function were added to the images with randomly chosen positions and magnitudes between 18 $\leq$ m$_{475}$ $\leq$ 31 and $-$0.75 $\leq$ (m$_{475}-$m$_{814}$) $\leq$ 3.50 and the images were re-photometered in identical fashion to the originals. The completeness is above 50\% down to limits of (m$_{475}$, m$_{814}$) = (28.2, 27.9), reaching the oldest MSTO with high reliability.  At these magnitude levels the typical photometric error is $\lesssim$0.1~mag, allowing us to resolve the old stellar sequences and make comparisons to isochrone models with a high degree of precision.

\begin{figure}[t!]
\plotone{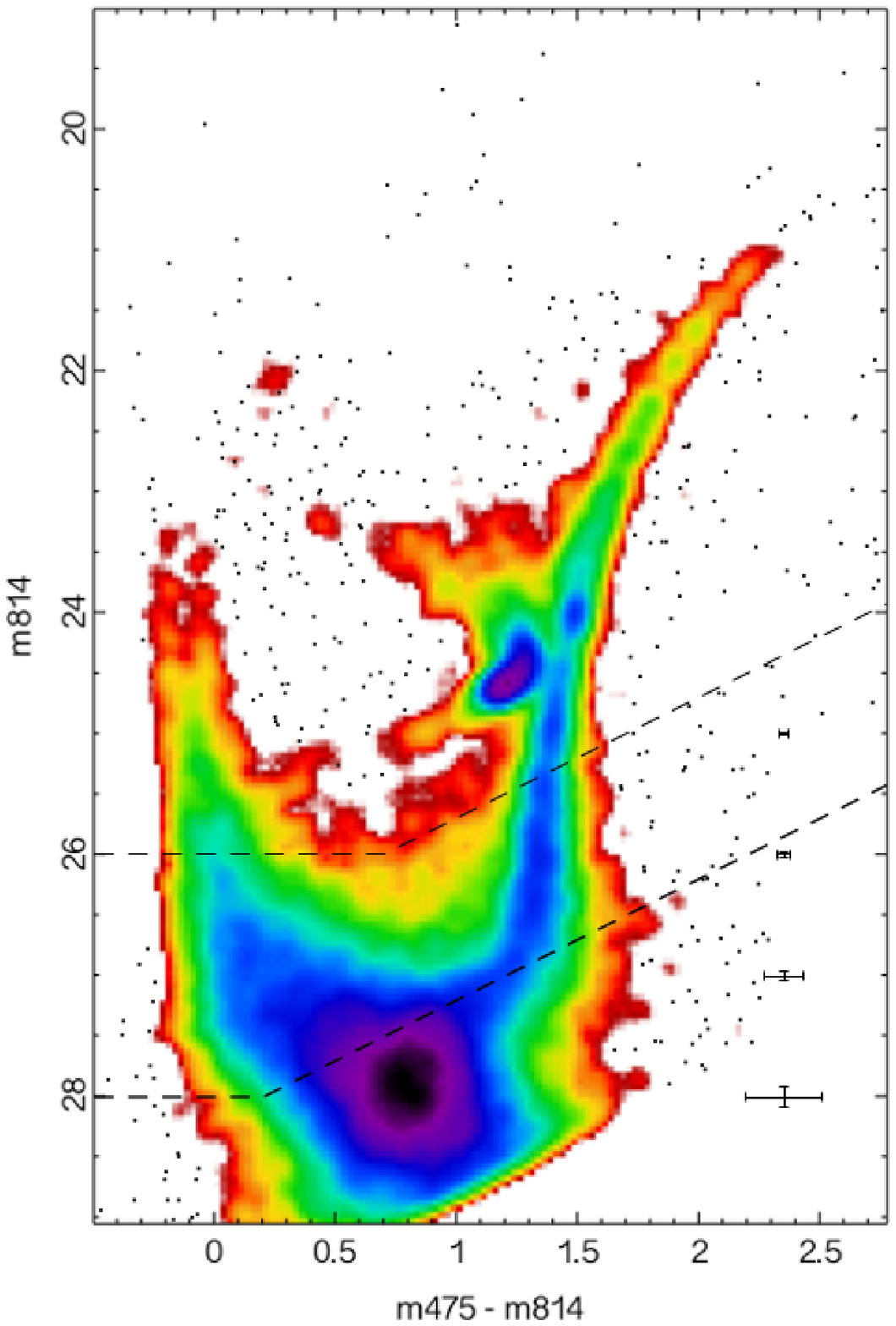}
\caption{Our ACS/WFC Hess diagram for Aquarius. Individual stars are plotted where the 
density is less than 4 stars per decimag$^{2}$. The colormap is logarithmic in order 
 to show the overall density distribution as well as the structure in the well-populated
stellar sequences.  Representative errorbars are shown, as are the 50\% and 90\% completeness 
limits.
\label{fig-cmd}}
\end{figure}

The resulting CMD is shown as a Hess diagram in Figure~\ref{fig-cmd}. 
The CMD is dominated by a very narrow red giant branch that terminates at m$_{814}$ = 21 and a main sequence indicating stars with a wide range of ages.  
We planned our observations using the conservative 1071~kpc distance estimate of \citet{mcc12}, leading us to expect high completeness only down to the level of the oldest subgiants, but because Aquarius is in fact $\approx$0.2~mag closer than this, we were able to achieve 50\% completeness at the level of the $\approx$10~Gyr MSTO.

\section{Stellar Populations of the Aquarius Dwarf} \label{sec:pops}
The stellar populations of Aquarius have been studied several times over the past two decades,
with increasing depth. \citet{mar90} made the first CCD-based study of the galaxy and commented on its
extremely faint, low surface brightness nature. 
The most thorough study of the stellar populations of Aquarius up to now has been the study of \citet{mcc06}, who combined Subaru and VLT data to study the red clump/horizontal branch for the first time, discovering that the CMD of the upper giant branch and red clump implied a young overall age for the galaxy, with most of the stars younger than 4$^{+2}_{-1}$~Gyr and fewer than 30\% older than 10~Gyr.

Our CMD is the deepest ever presented for Aquarius by more than 3~magnitudes, opening a unique window on the intermediate-age (1--10~Gyr) and old ($\geq$10~Gyr) stellar populations.
Major features in the luminosity function of the red giant branch are the very well defined red clump centered at m$_{814}$ = 24.6 and the RGB bump at m$_{814}$ = 24.  The well-defined RGB bump located above the red clump indicates that the dominant stellar population has a mean age quite a bit younger than the Milky Way globular clusters, but not a very high metallicity \citep{mon10b}. There are very few bright AGB stars above the RGB tip, consistent with the very low number of carbon stars reported by \citet{bat00} and \citet{gul07}, and suggesting that the dominant population is older than 1--2~Gyr.

The sequence of core helium-burning stars begins above the upper main sequence at m$_{814}$ $\approx$21 and continues downward and to the right through the red clump, reversing to form an extended horizontal branch. The continuity of these sequences shows immediately that while stars of all 
 ages from ancient (blue horizontal branch) to $\lesssim$10$^8$ yr (blue supergiants) are present, Aquarius is dominated by a population tending to the old side of intermediate age \citep[red clump/red horizontal branch;][]{col99a}.  

Overall, the impression given by the CMD is of a sparsely populated, but in many ways typical, dwarf irregular that has formed stars more or less continuously over a Hubble time. The upper main sequence, brighter than m$_{814}$ $\approx$ 24, is sparsely sampled due to the low star formation rate and stochastic sampling of the initial mass function for stars of the corresponding intermediate and high masses. Based on the low number of main sequence stars brighter than m$_{814}$ $\approx$22, Aquarius is experiencing a lull in star formation at present despite retaining enough gas to continue forming stars at a significant rate for several more Gyr \citep[e.g.,][]{hun85}.

A significant discovery in the context of this project is the blue horizontal branch (BHB), detected as an extension of the red horizontal branch to colours m$_{814-475}$ $\lesssim$0.5, overlapping the main sequence at magnitude m$_{814}$ $\approx 25.7$. These stars are unique signatures of an ancient, metal-poor stellar population, comparable to those seen in the classic satellite dwarf spheroidals of the Milky Way, for example the Sculptor dSph \citep{sal13}, as well as in virtually all dwarf irregulars imaged to sufficient depth \citep[e.g., IC1613,][]{col99b}. By contrast, Leo~A was found to have virtually no BHB stars of its own \citep{col07}, despite its similarity to Aquarius in both {\it present-day} stellar \citep*{mcc12} and dynamical \citep{beg06,kir14} masses.

RR~Lyrae type variables are identified in the HST observations, and these stars represent a
valuable independent probe of the distance and the metallicity of the oldest stellar populations. An analysis of the variable star populations, including the RR~Lyraes, will be presented in a forthcoming paper.
Because the fraction of BHB stars does not trivially scale to relative SFR at ages $\approx$10~Gyr or older \citep[e.g.,][]{sal13}, the degree to which the intermediate-age population is dominant, and the age and metallicity of the dominant populations, cannot be determined from counts of stars on the BHB alone.

\section{Star Formation History} \label{sec:sfh}
\subsection{Methodology}

\begin{figure*}[t!]
\plotone{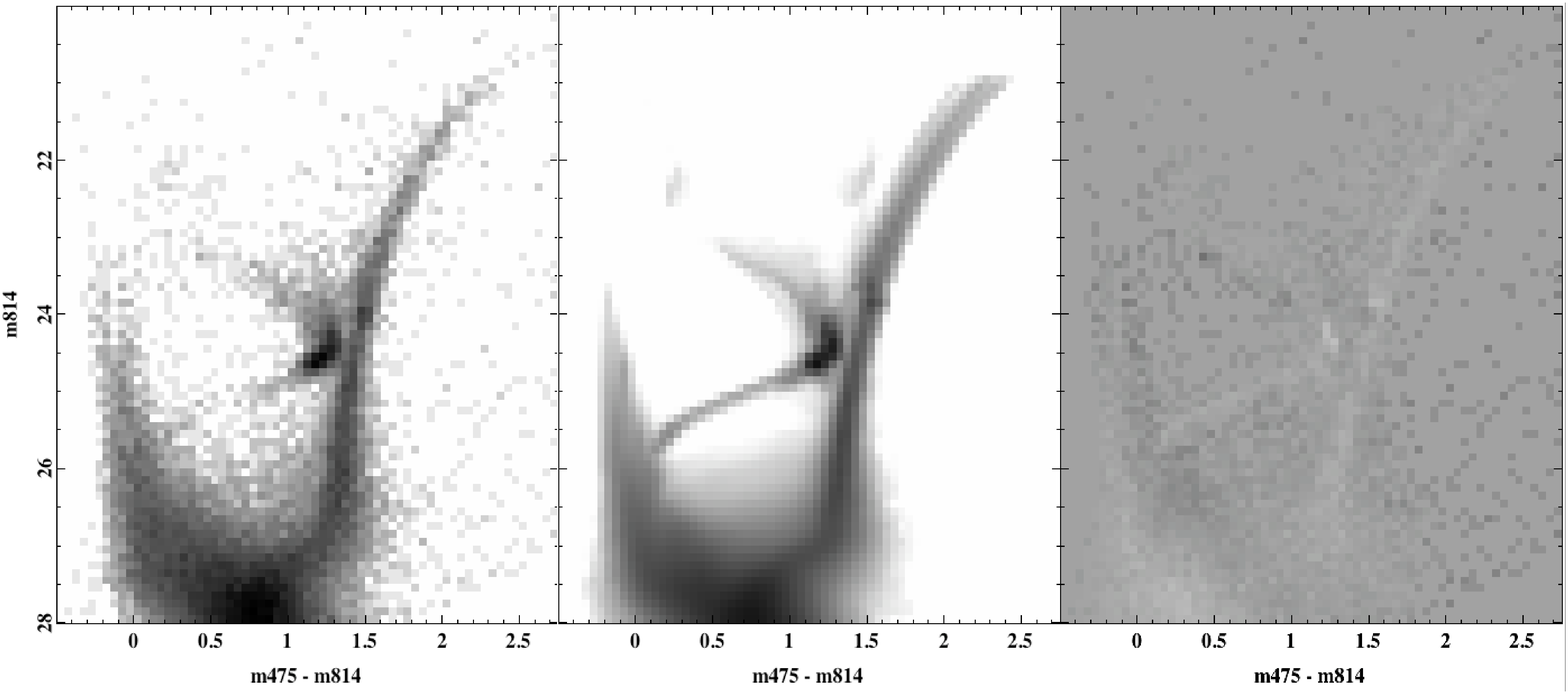}
\caption{Observed (left) and best-fit Hess diagrams of Aquarius showing the solution returned by synthetic CMD analysis using the PARSEC isochrone set \citep{bre12} and software written by Cole. {\it Left:} data. {\it Center:} model. {\it Right:} Residual significance, [(data$-$model)/data]$^{1/2}$.
\label{fig-hessfit}}
\end{figure*}

The SFH presented here has been calculated using the code developed by Cole, which has been used extensively on other Local Group dwarfs and tested against other methods with good consistency \citep[e.g.,][]{ski03,col07,mon10a,cig12,lia13,ski14}. Here, we use the Padova-Trieste \citep[PARSEC;][]{bre12} isochrones. These are  preferred because they have the most up to date input physics, include post-helium-core-flash phases of evolution, cover a wide range of age and metallicity at user-definable grid spacings, and are tuned to the revised solar oxygen abundance  \citep*[see, e.g.,][]{bah05}.  This specific choice plays a role in determining the precise age and duration of episodes of high or low SFR and the overall fit quality of the synthetic CMD, but the broad conclusions are robust against the choice of isochrone library owing to the depth of the photometry \citep{wei11}.

The initial mass function is that of \citet{cha03}, which is a Salpeter-like power law for M/M$_{\odot}$ $\geq$ 0.8 and log-normal for lower masses.  Binary stars are included as per the statistics presented in \citet{duq91} and \cite{maz92}, whereby 35\% of the ``stars'' are single objects, and the rest are binary; in turn, 25\% of the binaries are defined as ``close'' binaries in which the secondary mass is drawn from a flat IMF instead of the Chabrier IMF. The choice of IMF and binary statistics may slightly alter the mass normalization of the solutions but do not significantly affect the shape of the derived SFH, as discussed extensively in \citet{mon10b} and \citet{cig12}, among others. The distance and reddening were initially set to the values derived by \citet{mcc06}, but were allowed to vary as demanded by the fitting procedure in order to best match the observed CMD. The stability of the solutions has been confirmed by comparison to fits made independently with different choice of IMF and binary fraction, using the MATCH software package \citep{dol02}.

The isochrone library contains stars of ages between 6.60 $\leq$ log(Age/yr) $\leq$ 10.12, spaced by intervals of 0.01, and 
metallicities from 1/150th to 1/10 solar, evenly spaced at intervals of $\approx$0.15 in log(Z).
Multiple metallicities are allowed at a given age and we do not impose an age-metallicity relation (AMR; we do exclude metallicities above one-tenth solar from consideration).  The isochrones at each metallicity are grouped into age bins in order to speed the computation;
the age bins are initially taken to be evenly spaced by 0.10 in log(age), but the bins may grow or shrink depending on the noise level in the solutions that may justify finer or coarser spacing.

We adopted color and magnitude bins 0.05 and 0.10 mag wide, respectively. Because the data quality is high, we model the region with $-0.5 \leq (m_{814}-m_{475}) \leq 2.75$ and $19 \leq m_{814} \leq 29$, but we have checked that our conclusions are not significantly altered if these limits are shifted to brighter magnitudes. The fitting procedure accepted the adopted reddening value of E(B$-$V) = 0.05 and the \citet{tul09} distance modulus
(m$-$M)$_0$ = 24.95, but a good fit could not be found using the longer distance modulus from \citet{mcc12}. As a check we performed a TRGB fit and found good consistency for distances $\approx$977$\pm$45~kpc; the distance issue will be explored further in our paper presenting the variable star light curves (Skillman et al., in preparation).

The best-fit synthetic CMD is shown alongside the binned data in Figure~\ref{fig-hessfit}. The overall fit quality is very high, as shown by the residual significance plot in the righthand panel. The RGB color and breadth, the magnitude, color and morphology of the red clump, the upper main sequence color, and the location and slope of the SGB are all well-reproduced. The locations of the blue loop and red supergiant stars are indistinct due to the small number statistics and foreground contamination, but appears to be reasonable for metallicities Z $\approx$ Z$_{\sun}$/20.  The independent fits made using MATCH  gave very similar solutions within the uncertainties. 

\subsection{The Age of Aquarius}
\begin{figure}[t!]
\plotone{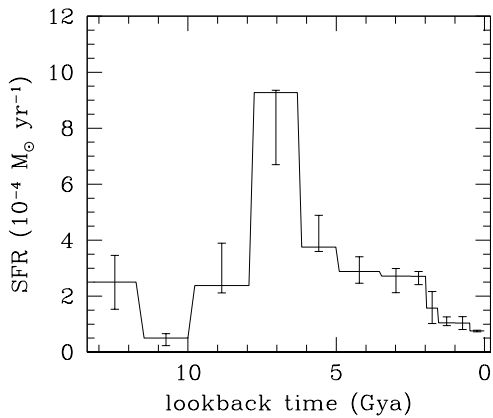}
\caption{The maximum likelihood star formation history of Aquarius, derived using the 
PARSEC isochrone set \citep{bre12} using software written by Cole. The dominant intermediate-age burst is robust against changes to the model details including choice of isochrone set and distance; see text for details. The errorbars include both random and systematic errors. For clarity the SFH younger than 1~Gyr has been combined into two bins.
\label{fig-sfh}}
\end{figure}

We present the lifetime SFH of Aquarius in Figure~\ref{fig-sfh}.  The major episode of star formation in Aquarius took place $\approx$6--8~Gyr ago, although star formation has been continuous (on Gyr timescales) over its entire lifetime. Although star formation at the earliest times is low, it is inconsistent with zero.  However, the drop in SFR between ages of 10--12~Gyr is robust and significant; the SFR in this age bin is only nonzero at the $\approx$2$\sigma$ level.
After the burst of activity at intermediate ages the SFR dropped, and has declined further over the past $\approx$2~Gyr. 

Aquarius is one of the least luminous gas-rich galaxies in the Local Group, but it does extend beyond the footprint of the ACS/WFC field (202$^{\prime\prime}$). \citet{mcc06} trace the structure out to at least a radius of 4$^{\prime}$, so our field only covers $\approx$50\% of the area over which Aquarius stars are known. This is potentially problematic for interpreting the age of Aquarius based on our SFH, given the ubiquity of age gradients and extended structures in even relatively isolated irregular galaxies-- for example, the case of Sextans~A and B \citep{doh97,bel14}. It has already been noted that the young stars in Aquarius are slightly off-center from the \ion{H}{1} contours \citep{mcc06}, which may be suggestive of a possible recent disturbance.  

However, there is good reason to believe that our SFH is giving a nearly complete picture of the history of Aquarius. The surface density profiles in \citet{mcc06} drop exponentially, becoming indistinguishable from the foreground population at $\approx$4$\arcmin$ along the major axis; the equivalent distance along the minor axis is 2$\arcmin$. 
\citet{mcc06} give an exponential fit to the elliptical surface density profile of Aquarius; by integrating their profile over the area of the ACS/WFC footprint, we find that 80\% of the Aquarius members brighter than V = 26 are contained in our field.
The integral of our SFR(t) over the Hubble time gives a total astrated mass $\approx4 \times10^{6}$ M$_{\odot}$; based on the profiles from \citet{mcc06} we may be missing $\approx10^{6}$ M$_{\odot}$.  Even if {\it all} of the missed stars were formed prior to 10~Gyr ago, then at the very most the derived full-galaxy SFH would show a time-averaged SFR from 10--13.7~Gyr of $\lesssim$40\% of the measured SFR from 6--8~Gyr ago. Even in this unrealistic ``oldest case'' scenario for the stars in the outskirts of Aquarius, the galaxy as a whole would be dominated by intermediate age stars.

As a further check for possible population gradients, and to attempt to characterize the outer structure of Aquarius, we obtained parallel imaging with the HST Wide Field Camera 3 (WFC3) simultaneously with our ACS images of the central field. 
The WFC3 field is centered $\approx$5.8$^{\prime}$ (1.65~kpc) from the center of the ACS/WFC field, nearly along the major axis. Despite a high degree of completeness down to magnitude $\approx$28, we were unable to identify any sign of the stellar sequences in Aquarius in these images.   At most a few Aquarius stars per square arcminute could be in the WFC3 field; unless the stellar distribution is highly asymmetric and we have been very unlucky in our parallel field placement, no significant population of Aquarius stars is present at radii $\gtrsim$5$^{\prime}$ (1.4~kpc), down to very strict limits around the old main sequence turnoff.

\begin{figure}[t!]
\plotone{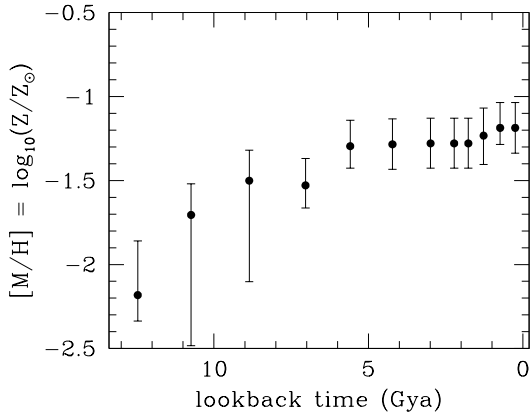}
\caption{The maximum likelihood age-metallicity relation of Aquarius, derived using the PARSEC isochrone set \citep{bre12}. This scale assumes a solar heavy element mass fraction Z$_{\odot}$ = 0.015. A range of metallicities is present at most ages; the errorbars show the rms metallicity range of stars at a given age.
\label{fig-amr}}
\end{figure}

\subsection{Age-Metallicity Relation}
The metallicity evolution of Aquarius is summarized in Figure~\ref{fig-amr}, and is in excellent agreement with the spectroscopic metallicities reported by \citet{kir14} and with the photometric estimates in \citet{mcc06}. The errorbars in Fig.~\ref{fig-amr} show the rms of the mass-weighted metallicity in each age bin; because the isochrone models are separated by discrete intervals of $\approx$0.15--0.20 log metallicity, the errorbars are typically about this size, except where a very wide range of metallicities are required to reproduce the CMD, as in the oldest age bins. The wide metallicity range for ages $\gtrsim$8~Gyr likely reflects both the rapid early increase in metallicity and the small number of stars in the CMD at those ages. Because of the rather coarse age resolution at the earliest times, we are unable to unambiguously assign an age to the youngest major populations with Z $<$ 0.01Z$_{\sun}$.  After the period of rapid early enrichment, ubiquitous in synthetic CMD solutions for dwarf galaxies, the metallicity remains nearly constant over time, slowly rising to a value of $\approx$Z$_{\odot}$/20, fairly typical for dwarf galaxies of this luminosity \citep[e.g.,][]{van97}. The mean metallicity at intermediate ages is in good agreement with the Local Group dwarf galaxy stellar mass-metallicity relationships reported by \citet{ber12} and by \citet{kir13}.

\subsection{Recent SFH}
\begin{figure}[t!]
\plotone{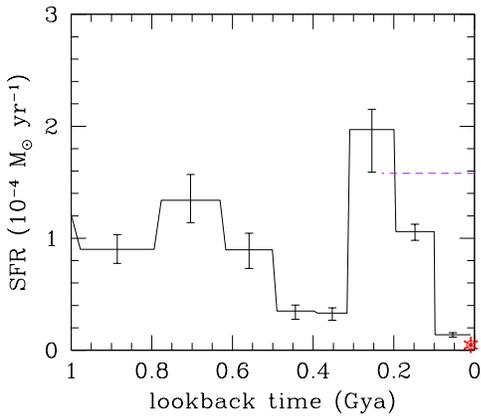}
\caption{A detailed look at the SFH of Aquarius over the past Gyr. The SFR appears to be dominated by stochastic fluctuations with a very low mean level. The enhancement at $\approx$250--300~Myr is easily visible in the CMD.
The dashed line indicates the SFR derived from GALEX observations, and the circled star gives the H$\alpha$ SFR \citep{lee09}.
\label{fig-recent}}
\end{figure}
\begin{figure*}[t!]
\plotone{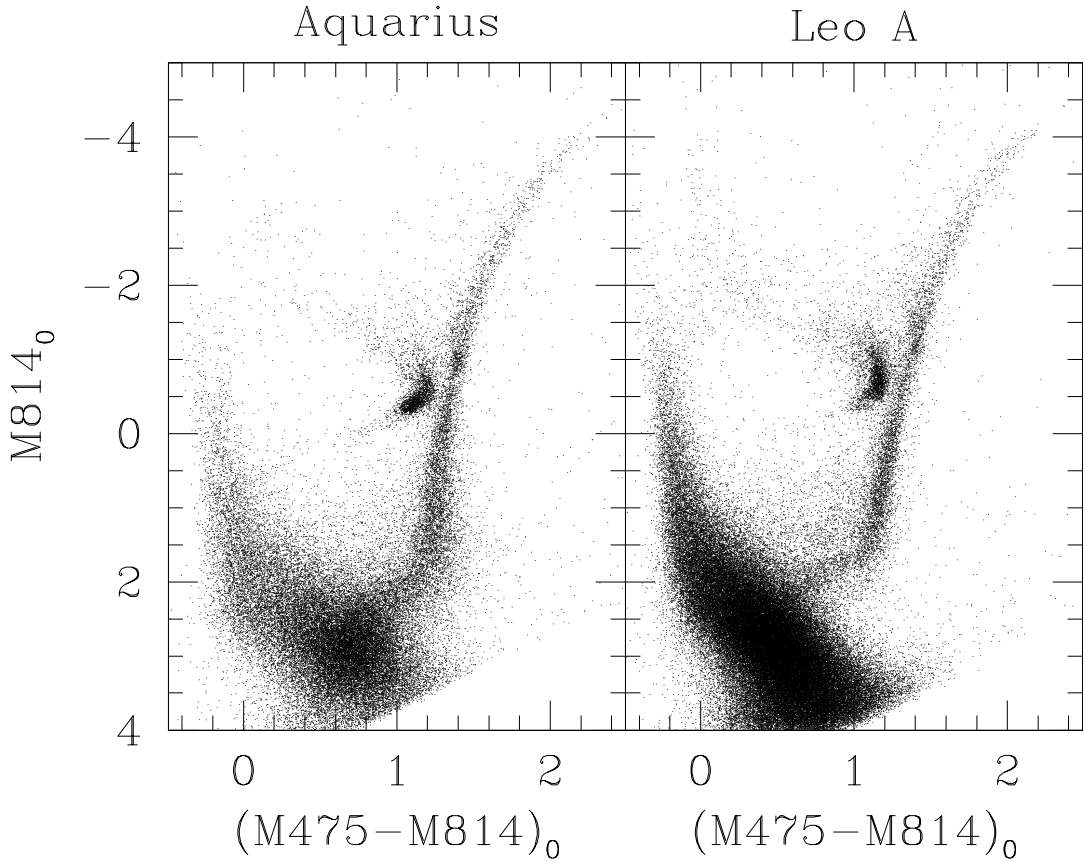}
\caption{Our CMD of Aquarius, corrected for distance and reddening, beside the analogous
CMD of Leo~A \citep{col07}. Differences in the red clump/HB morphology and the extent of the subgiant branch make plain that Leo~A and Aquarius do not share a common quenching time. While both galaxies contain genuinely ancient stars, the relative fraction of old stars is over twice as large in Aquarius.
\label{fig-compare}}
\end{figure*}

Narrowband imaging and optical spectroscopy at H$\alpha$ by \citet{van97} revealed no current massive star formation activity, even though the presence of bright blue main sequence and evolved stars proves that star formation has occurred within the past 10$^8$ yr.  
Aquarius is one of the faintest, lowest star formation rate (SFR) galaxies in the 11HUGS sample of gas-rich, star-forming galaxies within 11~Mpc \citep{ken08}, with a large discrepancy between the inferred H$\alpha$ and far-ultraviolet SFR owing to the flickering nature of star formation in dwarf irregulars, combined with stochastic sampling of the initial mass function (IMF) at low SFR \citep*{lee09,fum11,wei12b}. 

A higher resolution view of the SFH over the past 1~Gyr is shown in Figure~\ref{fig-recent}. Aquarius is in a period of low SFR, with seemingly random fluctuations on timescales of $\approx$10$^8$~yr.  The recent drop in activity after an upward fluctuation is in excellent agreement with the estimates of current SFR reported in \citet{lee09}, whose UV ($\sim$100~Myr sensitivity; dashed line) and H$\alpha$ ($\sim$10~Myr sensitivity; circled star) results are overplotted on Figure~\ref{fig-recent}. The enhancement at ages of 250--300~Myr has been noted by previous authors as discussed above, and is identifiable here by the increased density of blue loop stars at m$_{814}$ $\approx$22. It is natural to interpret this increase in SFR lasting several tens of Myr as a stochastic fluctuation in SFR of the type frequently observed in dwarf galaxies, leading to the wide variety of H$\alpha$, UV, and IR star formation indicators in dwarf galaxies where the gas densities and total mass participating in star formation are small \citep[e.g.,][]{lee09}. The event is of shorter duration than a typical episode that would be identified as a dwarf galaxy starburst \citep{mcq10}.

\section{Aquarius in Context with Other Dwarf Galaxies} \label{sec:disc}

The isolation of Aquarius means that it has never been affected by tidal encounters with the Milky Way or M31, and so it is highly unlikely that it was of much higher mass in the distant past. This removes one complicating factor in the interpretation of the SFHs of close satellites, which may have been of much higher mass in the distant past than they are now \citep*[e.g.,][]{pen08}. On the contrary, a field dwarf like Aquarius is likely to have only accreted about half of its present-day mass prior to $\sim$10~Gyr ago \citep[e.g.,][]{bro14}. 

In making comparisons to other galaxies of the Local Group, it is crucial to remember that systematic errors in the SFH at old ages are unavoidable when shallower CMDs are used \citep{wei11}, and the most meaningful comparisons are to be made comparing CMDs of similar absolute depth \citep[e.g.,][]{wei14a}.  This tends to limit the gold standard sample of isolated comparison galaxies to the LCID targets \citep{hid13} and Leo~T \citep{wei12} but still allows some patterns to emerge. Grouped by overall history of star formation, the galaxy most similar to Aquarius is the late-blooming dIrr Leo~A \citep{col07}, which although similarly isolated and of similar dynamical mass was delayed for longer and formed stars much more efficiently after igniting at intermediate age.

\subsection{Leo~A (DDO 69)}

\begin{figure*}[t!]
\plotone{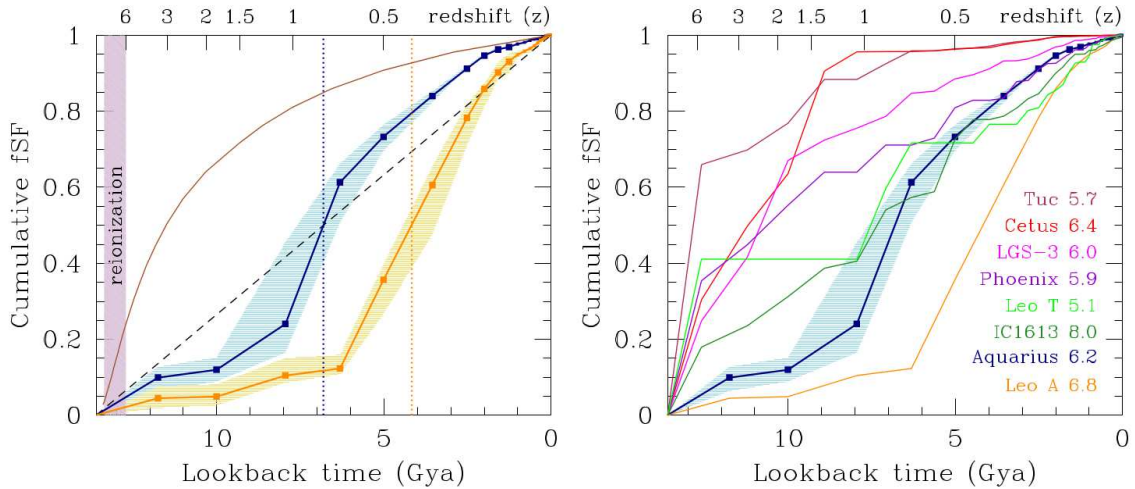}
\caption{The cumulative fractional star formation of Aquarius (blue) and Leo~A (orange) compared to each other and to other Local Group galaxies.  The redshift scale at the top is calculated assuming a concordance $\Lambda$CDM cosmology. 
{\it Left:} The median ages of the two galaxies are indicated by vertical dotted lines. The dashed line shows a constant SFR, while the convex-upward solid curve shows a reference dark matter assembly history for a 10$^{10}$ M$_{\sun}$ 
simulated galaxy \citep{fak10}. The shaded portion of the graph shows the era of cosmological reionization, from $z \approx$6--14. {\it Right:} the cumulative SFH of Aquarius compared to seven other galaxies with comparably precise SFH data
(see text for details). The estimated stellar mass \citep[log M/M$_{\sun}$;][]{mcc12} is given for each galaxy. Aquarius and Leo~A are the only two galaxies in the sample of eight to show SFRs significantly below their lifetime average for the first few Gyr after reionization.  
\label{fig-thepoint}}
\end{figure*}

In Leo~A, the next most-isolated galaxy to be studied in equivalent detail, star formation also
began at the oldest ages, but continued at only a very low level until $\approx$6~Gyr ago
\citep{col07}.  Like Aquarius, Leo~A has almost certainly never interacted with another (known) galaxy \citep[e.g.,][]{sha13}. The two galaxies are similar in their degree of isolation, total magnitude M$_B$, color, and gas content, so any similarities that can be identified in their SFH are of great interest.

Comparing the dynamical masses within the half-light radii, \citet{kir14} gives M$_{1/2}$ = 15$^{+6}_{-5}\times10^{6}$~M$_{\odot}$ for Leo~A and 20$^{+10}_{-8}\times10^{6}$~M$_{\odot}$ for Aquarius; the latter value may be reduced by 9\% to account for the revised distance estimate presented in this paper (see Table~\ref{tab:props}). 
With that correction the two galaxies are seen to have nearly the same total mass, although they have somewhat different 
mass to light ratios, in part because of the large number of young stars in Leo~A. Leo~A also supports a higher current SFR, has a higher gas fraction \citep{mcc12} and a slightly lower present-day metallicity \citep{kir13}.

In both galaxies, the total stellar mass reported by different authors varies based on the adopted distance and IMF, and details of the integrated light photometry. Most reported stellar mass estimates for Leo~A ultimately derive either from the work of \citet{lee03} or \citet{van04}; correcting these to the consistent distances derived in \citet{dol02b} and \citet{col07} we conclude that the best estimate of the stellar mass of Leo~A is $\approx$3$\pm1\times$10$^6$ M$_{\sun}$. Fewer measurements are available for Aquarius; using the method of \cite{lee03} to convert from integrated magnitude and B$-$V color to stellar mass indicates that Aquarius has $\approx$1/2 \citep[using the photometry of][]{lee99} to $\approx$1/3 
\citep[using the photometry of][]{mcc06} of the stellar mass of Leo~A, $\approx$1--2$\times$10$^6$ M$_{\sun}$.

The CMDs of Aquarius and Leo~A, corrected for distance and reddening, are shown side by side in 
Figure~\ref{fig-compare}, highlighting their overall similarity and differences in detail. Apart from the quite different HB/red clump morphology, the biggest difference is in the subgiant branches (SGB), where Leo~A is much more strongly concentrated aound M814$_0$ $\approx$ $+$1.5--2, with only a scattering of fainter stars. 

In order to quantify the similarities and differences in the SFH of Aquarius and Leo~A, we rederive the Leo~A SFH using the data from \citet{col07} with the identical stellar evolution library and time bins used in our solution for Aquarius. The results are entirely consistent with the SFH reported in \citet{col07}; the two galaxies are directly compared in Figure~\ref{fig-thepoint}.
It is seen that the drought in star formation was broken earlier in Aquarius, but that the galaxy then rapidly dropped back down to nearly its lifetime average SFR. By contrast, Leo~A formed stars at a very slow rate over much of its lifetime and only managed to start converting gas into stars in earnest $\approx$2~Gyr {\it after} Aquarius. The median age of star formation (the time by which 50\% of the mass of all stars ever formed had been created) for each galaxy is shown with the vertical dotted lines; the median age of Aquarius is 6.8~Gyr, but for Leo~A it is 4.2~Gyr.

\begin{deluxetable*}{cc|ccc|ccc}[h!]
 \tablecolumns{8}
 \tablewidth{\textwidth} 
 \tablecaption{Comparative SFH of Aquarius \& Leo~A\label{tab:compare}}
 \tablehead{
   \colhead{Lookback Time} &
   \colhead{Redshift\tablenotemark{a}} &
   \multicolumn{3}{c}{Aquarius} &
   \multicolumn{3}{c}{Leo A} \\
   \cline{3-5} \cline{6-8} \\
   \colhead{(Gya)} & 
   \colhead{($z$)} & 
   \multicolumn{1}{|c}{SFR\tablenotemark{b} ($10^{-4}$ M$_{\odot}\cdot$yr$^{-1}$)} &
   \colhead{$\int$M$_{\star}$\tablenotemark{c} ($10^6$ M$_{\odot}$)} &
   \multicolumn{1}{c|}{Z/Z$_{\sun}$\tablenotemark{d}} &  
   \multicolumn{1}{|c}{SFR ($10^{-4}$ M$_{\odot}\cdot$yr$^{-1}$)} &
   \colhead{$\int$M$_{\star}$ ($10^6$ M$_{\odot}$)} &
   \multicolumn{1}{c}{Z/Z$_{\sun}$}}
 \startdata
  0--0.5    & 0--0.04        & 0.76 $^{+0.06}_{-0.06}$  & 4.03 & 0.065 & 1.82 $^{+0.10}_{-0.10}$ & 3.54 & 0.051 \\[4pt]
  0.5--1    & 0.04--0.08   & 1.04 $^{+0.23}_{-0.23}$  & 3.99 & 0.065 & 1.73 $^{+0.14}_{-0.15}$ & 3.45 & 0.051 \\[4pt]
  1--2       & 0.08--0.2     & 1.26 $^{+0.44}_{-0.37}$  & 3.94 & 0.056 & 3.19 $^{+0.23}_{-0.20}$ & 3.36 & 0.041 \\[4pt]
  2--4       & 0.2--0.4       & 2.77 $^{+0.27}_{-0.44}$  & 3.81 & 0.053 & 5.70 $^{+0.53}_{-0.55}$ & 3.05 & 0.039 \\[4pt]
  4--6       & 0.4--0.7       & 3.28 $^{+0.88}_{-0.27}$  & 3.26 & 0.052 & 6.31 $^{+0.51}_{-0.80}$ & 1.91 & 0.039 \\[4pt]
  6--8       & 0.7--1.1       & 8.26 $^{+0.39}_{-1.87}$  & 2.60 & 0.034 & 1.33 $^{+0.14}_{-0.69}$ & 0.64 & 0.022 \\[4pt]
  8--10      & 1.1--1.8      & 2.38 $^{+1.51}_{-0.26}$  & 0.95 & 0.032 & 0.99 $^{+0.30}_{-0.10}$ & 0.38 & 0.026 \\[4pt]
  10--11.5 & 1.8--3         & 0.50 $^{+0.16}_{-0.27}$  & 0.48 & 0.020 & 0.12 $^{+0.26}_{-0.10}$ & 0.18 & 0.013 \\[4pt]
  $\geq$11.5 & $\geq$3 & 2.22 $^{+0.96}_{-0.97}$  & 0.40 & 0.007 & 0.90 $^{+0.22}_{-0.64}$ & 0.16 & 0.007
 \enddata
 \tablenotetext{a}{Assuming $\Omega$ = 1, $\Omega_{\Lambda}$ = 0.73, H$_0$ = 71 km$\cdot$s$^{-1}\cdot$Mpc$^{-1}$; \citet{wri06}.}
 \tablenotetext{b}{Average star formation rate over the given time period.}
 \tablenotetext{c}{$\int$SFR($t$)$dt$ from formation until the end of the given time period (excludes losses due to finite stellar lifetimes).}
 \tablenotetext{d}{Average metallicity (mass fraction of elements heavier than helium) of stars formed during the given time period, compared to the solar metallicity Z$_{\sun}$ = 0.0152.}
\end{deluxetable*}

The absolute star formation rates, the integral of SFR over lifetime, and metallicity evolution of the two galaxies are compared in Table~\ref{tab:compare}. 
The two galaxies are clearly of a very similar nature, with a nearly equal total astrated mass. It is interesting that, with the two solutions run completely independently and with no enforcement of a chemical evolution law, Aquarius is found to be more metal-rich than Leo~A at almost every age.  For the intermediate age stars that dominate the red giant sample of \citet{kir13}, the metallicity difference between the two galaxies is of the same magnitude and in the same sense as observed, with Leo~A about 0.1--0.2~dex more metal-poor. While Leo~A is $\approx$1~mag brighter than Aquarius at the present day, our modeling finds that its total lifetime astrated mass has been less than Aquarius; this difference is also in the same sense, and of the same order of magnitude as the dynamical mass difference reported by \citet{kir14}. Presumably the difference in metallicity may be related back to the more gas-rich nature of Leo~A, which is a factor of 2--3 more gas rich than Aquarius \citep{mcc12} and therefore less chemically evolved.  At the oldest ages, Aquarius did not just form stars at a higher relative rate than Leo~A, but had a higher absolute star formation rate (a fact that is reflected in its much more prominent blue horizontal branch). 

\subsection{Other Local Group Galaxies}

All other galaxies with comparably deep CMDs (in absolute magnitude) are much closer to the Milky Way\footnote{or in a few cases, M31.} and therefore complicated by the possibility that past interactions have strongly influenced their SFH.
\citet{mcc12} shows that the irregular and transition type galaxies IC~1613, Leo~T, Phoenix, and LGS-3 are all at least a factor of two closer to a massive spiral than Aquarius. The isolated systems WLM and Sagittarius DIG have been surveyed with WFPC2 \citep[see][]{wei14a}, but not to a depth sufficient to resolve the oldest MSTO; the deepest ACS data for UGC~4879 reach the level of the red clump \citep{jac11}.

The non-satellite dwarf spheroidals Tucana and Cetus are somewhat different. They differ from the gas-rich galaxies in their star formation histories; this may or may not be related to a proposed history of tidal interactions 
\citep[][for Tucana and Cetus respectively]{fra09,lew07}.
Nevertheless while interactions {\it may} be suspected for Tucana and Cetus \citep*[e.g.,][]{tey12}, the timing and identity of the main tidal perturber cannot be established with confidence.  They are included here for completeness.

Most of the Local Group dwarf galaxies show either nearly constant long-term average SFR, or are dominated by much older bursts. The latter pattern is more common in the nearby satellites of the Milky Way and M31 and among the lowest-luminosity systems. The Tucana and Cetus dwarf spheroidals are nearly pure ancient populations \citep[albeit with ages differing by $\approx$1~Gyr;][]{mon10a,mon10b}. The dwarf irregular galaxies, which statistically tend to be found at large distances from both M31 and the Milky Way, show more uniform SFRs over cosmic time. The prototypical example with comparable ACS data could be IC~1613, a far larger galaxy (M$_V$ = $-$15.2) which shows nearly constant SFR \citep{ski14}. At the other end of the mass scale, Leo~T (M$_V$ = $-$8) is observed to show a slowly declining SFR beginning at the earliest times and continuing until $\approx$25~Myr ago without dramatic variation \citep{wei12,cle12}.

From Figure~\ref{fig-thepoint}, we see that there is a wide range of observed SFHs at a given galaxy mass.  These galaxies are not satellites of either the Milky Way or M31, so no tidal or ram pressure quenching of star formation is expected, and indeed most of the systems show extended star forming lifetimes. Among the galaxies with a freefall time onto either the Milky Way or M31 less than a Hubble time, no examples of significantly delayed peak SFR are found. Leo~A and Aquarius, the two most isolated galaxies in the sample, on the other hand, simmer for at least 3--4~Gyr after reionization before dramatically increasing their SFR. Present-day isolation cannot be the sole factor determining the early SFH, however, because the similarly distant and similarly massive dwarf spheroidals, Cetus and Tucana, have the {\it highest} fraction of stars older than 10~Gyr among the galaxies in our sample.

The comparison to other Local Group galaxies serves to reinforce the point that there is no simple relationship between current galaxy mass, isolation, and age of the main stellar population; it is physically intuitive to suppose that such a relationship would be clarified if the complete history of interaction and feedback heating for each galaxy could be known. If there are any trends, it seems that major star formation events are more evenly distributed in redshift than in lookback time, and galaxies with low baryonic mass are more prone to bursty behavior than more massive and/or gas-rich systems.  

\begin{figure*}[t!]
\plotone{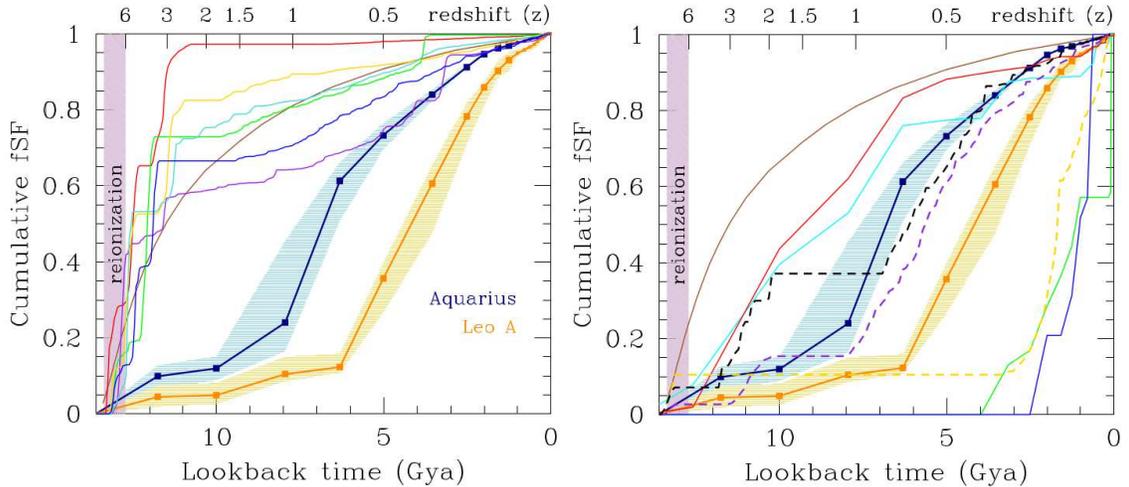}
\caption{The cumulative fractional star formation of Aquarius (blue) and Leo~A (orange), compared to cosmological simulations including baryons. {\it Left:} simulated dwarfs from \citet{saw11} for galaxies with similar halo mass to Aquarius and Leo~A; the SFH tracks the dark matter assembly and forms too many stars too early. {\it Right:} simulated dwarfs from \citet{she13} (solid lines, color-coded following their Fig.~5) and from \citet{bro14} (dashed lines). In both panels, the dark matter-only simulated 10$^{10}$~M$_{\sun}$ halo of \citet{fak10} is shown as the smooth curve.
\label{fig-models}}
\end{figure*}

\section{Implications for Galaxy Formation Models} \label{sec:models}

A significant amount of recent progress has been made in cosmological structure formation models describing the formation and early evolution of dwarf galaxies. The gains in numerical resolution and treatment of various radiative and hydrodynamical effects are reflected in simulated galaxies that plausibly resemble observed dwarfs. 

In the most thoroughly studied galaxies, their proximity means that environmental effects arising from close passages to the Milky Way must be considered when interpreting the SFH. This has weakened inferences about the effects of the UV background or internal feedback on the SFH of dwarfs. While the hypothesized effects are expected to be strongest for the lowest-mass galaxies, most of the extremely low-luminosity (``ultrafaint'') dwarfs are deep within the Milky Way potential well and therefore present an undesirable degree of complexity despite being the obvious place to test for feedback or photoionization heating. 
Aquarius and Leo~A have certainly never had a close flyby of either M31 or the Milky Way, and have likely never been within the virial radius of either galaxy \citep{sha13}; they are the two best ``clean'' galaxies we have for comparison to model predictions. 

\citet*{fak10} trace the assembly of dark matter halos beginning with 10$^6$~M$_{\odot}$ clumps at redshift $z$ = 15 and find that the merger rate is approximately constant {\it per unit redshift}, which means that typical dwarfs assembled most of their mass at very high lookback time. Their examination of the Millennium simulation finds that the median redshift of last major merger for 10$^{10}$~M$_{\sun}$ halos is $z$ = 3.4 (t$_{\mathrm{lookback}}$ = 11.8~Gyr). The SFH derived assuming that SFR follows accretion rate for a typical halo from \citet{fak10} is shown as the smooth curve in Figure~\ref{fig-thepoint}. 

While the median dwarf is expected to form early, \citet{fak10} also find a broad distribution of accretion rates at all redshifts for small halos. Their models predict that for halos with present-day virial mass 10$^{10}$~M$_{\sun}$, roughly 1 in 3  experienced their last major merger after $z$ = 2, and about 1 in 6 since $z$ = 1. The implication is that while late accretion is not the {\it most} likely evolutionary path for a dwarf galaxy the size of Aquarius, it is far from {\it un}likely. The fact that {\it both} of the first two isolated dwarfs to have been studied in sufficient detail show just such delays leads to the impression that either long delays are more common in isolated galaxies than in the galaxy population as a whole, or that star formation does not track dark matter accretion. If late mergers are the explanation for the delayed SFH, then neither component of a major merger could have had a dominant ancient stellar population.

While a simple prescription for baryon (re)heating after reionization might predict the complete quenching of star formation in dwarf galaxies at redshifts $z$ $\approx$6, there is little evidence in the SFHs of observed galaxies for such a quenching \citep{wei14b}. Rather, the effects of reionization, if any, appear to be more subtle; note in Figure~\ref{fig-thepoint} that no isolated Local Group dwarf begins vigorously forming stars within 1--2~Gyr after the end of reionization. While star formation does not appear to be quenched, an argument could be made that it is at least dampened by the UV photoheating: every galaxy in Figure~\ref{fig-thepoint} has a lower rate of star formation immediately after the end of reionization than before. 

Cast in the framework of hierarchical galaxy assembly in a $\Lambda$CDM cosmology, it appears likely that the accretion history of baryons in Aquarius either differs significantly from that of the dark matter, or baryonic physics has disrupted the one-to-one link between mass accretion and star formation. The observed complexity of these relationships has proven challenging for non-fully cosmological models to capture, and predictions for stellar and dark matter mass growth are not always in agreement with observations \citep[e.g.,][]{lei12,beh13,lu14a,lu14b,wei14b}.

The most recent generation of cosmological structure formation models features high numerical resolution and advanced prescriptions for feedback, star formation, variable UV background, and radiative cooling through both H$_2$ and metal lines   \citep[e.g.,][]{chr12,she13,hop13}. The effects of reionization and feedback are typically more subtle and varied in recent simulations than in previous work, and it appears that progress is being made on the overcooling front.

Figure~\ref{fig-models} compares the SFH of Aquarius and Leo~A to the predictions from three recent sets of simulations that highlight the problems associated with dwarf galaxy formation and the recent progress that has been made. The left panel shows six isolated model dwarfs from the hydrodynamical simulations of \citet{saw11}. The simulated galaxies have a dark matter halo mass of $\approx$10$^{10}$~M$_{\sun}$, comparable to or a bit larger than Aquarius. In the simulations, star formation is suppressed by the combined effects of feedback and reionization, but unlike Aquarius and Leo~A, the SFR never recovers. Despite an advanced cooling and star formation treatment, the simulations form the majority of their stars before $z$ $\approx$2 because SFR tracks the accretion rate. The simulated galaxies also form a minimum of an order of magnitude too much stellar mass compared to the observed dwarfs (5--10$\times$10$^{7}$~M$_{\sun}$; for comparison see Table~\ref{tab:compare}).

Even more recent models are shown in the right hand panel. The dashed lines show the three galaxies in the simulations of \citet{bro14} that have produced on the order of 10$^{6-7}$~M$_{\sun}$ worth of stars, similar to Aquarius. These simulated isolated dwarfs are gas-rich (M$_{HI}$/M$_{\star}$ $>$1), dark matter-dominated, and have experienced a bursty SFH. The most gas-rich, lowest stellar-mass dwarf is an obvious comparison object to Leo~A, with some initial star formation prior to the end of reionization, followed by a long fallow period and a late burst. The other two simulated dwarfs also show an initial epoch of star formation with a post-reionization drought, although star formation restarts after a much briefer delay. 
\citet{bro14} conclude that some {\it isolated} dwarfs are capable of retaining gas without forming stars for several Gyr. A complete analysis of the relationship between mass accretion, feedback, gas cooling, and SFR in these simulations will be presented in a forthcoming paper (Weisz et al., in preparation), but the general agreement is quite promising and may indicate that a critical threshold in resolution has been crossed. 

The solid lines in the right-hand panel of Figure~\ref{fig-models} show the results of \citet{she13}, whose high-resolution smooth-particle hydrodynamical model also shows that the combination of reionization and the mechanical energy input by supernova feedback is critical to the prevention of gas cooling in small halos. They make a strong prediction that halos with total mass $\lesssim$10$^9$~M$_{\odot}$ never manage to cool enough to form stars; this threshold is just below the expected virial masses of Aquarius and Leo~A.  Four of their simulated galaxies span the range of plausible masses for our isolated dwarfs. The nearly constant SFR dwarfs ``Bashful'' and ``Doc'' inhabit more massive halos of $\approx$1--3$\times$10$^{10}$~M$_{\sun}$. In Figure~\ref{fig-models}, these galaxies have formed the bulk of their stars far earlier than Aquarius or Leo~A, and have also formed over an order of magnitude more stellar mass that the isolated Local Group dwarfs.

The other two star-forming galaxies from \citet{she13}, ``Dopey'' and ``Grumpy'', are at the lower end of the applicable halo mass range, and present a stark contrast to the more massive systems: they are extremely gas-rich and have formed 1--2 orders of magnitude fewer stars than Aquarius or Leo~A. The low-mass simulated dwarfs exhibit a highly bursty SFH, and are characterized by an extreme delay of star formation-- neither galaxy experiences {\it any} star formation until very recently, making them exaggerated versions of Leo~A. The variance in SFH between any two given galaxies of similar total mass indicates a large degree of stochasticity in all three models considered here, a phenomenon which is also expected in the models of, for example, \citet{sta13} and others.

The fact that realistic small galaxies form at all in the simulations represents an encouraging sign of progress, but there are still significant discrepancies. A lingering sign of the overcooling problem is that the extrapolated \citet{beg06} NFW profile mass for Aquarius is $\approx$10$^{10}$, similar to the Shen's  ``Bashful'' and ``Doc'', but these produce ten times as much stellar mass as Aquarius and still have a very large amount of gas left over. Scaling the mass down by 0.5--1 order of magnitude for the other two simulated galaxies reduces the number of stars formed by 2--3 orders of magnitude, indicating that we are in a mass range where the feedback and star formation efficiency prescriptions are exceptionally sensitive to halo mass.  By contrast, the most and least massive galaxies in Figure~\ref{fig-compare} (IC~1613 and Leo~T, respectively) share broadly similar patterns of relative star formation rate, are both gas-rich at the present time, and are both older than the intermediate-mass galaxies Aquarius and Leo~A.

\section{Summary and Conclusions}
We have measured the star formation history of the Aquarius dwarf galaxy based on photometry extending to the oldest main-sequence turnoff obtained with the HST Advanced Camera for Surveys. We have fit the photometric data using the method of creating synthetic color-magnitude diagrams based on theoretical isochrones and a maximum-likelihood comparison. We have found that Aquarius is dominated by stars aged less than 10~Gyr, and that metallicity evolution over the past $\approx$10~Gyr has been mild.  We find in particular the following major features:

\begin{enumerate}
\item The SFH is not consistent with a strong, early burst of star formation prior to the end of reionization. However, a significant population of stars older than 10~Gyr is present based on the SFH, the magnitude extent of the subgiant branch, and the detection of a blue horizontal branch.

\item The first major episode of star formation began $\approx$8~Gyr ago and lasted for $\approx$2~Gyr. Star formation has lasted for the full age of the Universe, with no evidence for long breaks between active periods. Apart from the increase at 8~Gyr, the SFH is almost consistent with a constant lifetime SFR.  In this sense Aquarius resembles Leo~A, which is similarly isolated and of comparable mass, but experienced an even stronger delay. The presence of an old stellar population amounting to 4$\times$10$^5$~M$_{\odot}$ in place prior to redshift $z$ = 3 was evidently not accompanied by sufficient feedback to quench subsequent star formation.
 
\end{enumerate}

Aquarius is the most isolated galaxy for which the full SFH has been measured with high precision and accuracy based on photometry reaching the oldest MSTO; it therefore makes an ideal test case to benchmark the evolution of small galaxies in isolation. Either Aquarius experienced a ``slow drip'' accretion history, in which gas trickled in over several gigayears before reaching a threshhold density for star formation, or feedback effects on the gas have decoupled the star formation rate from the gas accretion rate.

The differences between Aquarius and Leo~A suggest that some combination of inefficient cooling in small halos, heating due to feedback, and/or a stochastic history of baryon accretion is necessary to understand the evolution of the smallest galaxies. Because of the high level of variance observed in the SFH of isolated dwarf galaxies to date, observations of additional galaxies to comparable photometric depth will be necessary in order to begin to map correlations in the space of total mass, baryon fraction, angular momentum, and environment. 
  
\acknowledgments 
We would like to thank Sijing Shen and Till Sawala for sharing the results of their simulations with us.
Support for DRW is provided by NASA through Hubble Fellowship grant HST-HF-51331.01 awarded by the Space Telescope Science Institute. AB, EDS, and DRW were supported in part by National Science Foundation Grant No.\ PHYS-1066293 and the hospitality of the Aspen Center for Physics. RL acknowledges financial support to the DAGAL network from the People Programme (Marie Curie Actions) of the European Union's Seventh Framework Programme FP7/2007--2013/ under REA grant agreement number PITN-GA-2011-289313. Additional support for this work was provided by NASA through grant number HST~GO-12925 from the Space Telescope Science Institute, which is operated by AURA, Inc., under NASA contract NAS5-26555. This research made extensive use of NASA's Astrophysics Data System bibliographic services.

\clearpage

\end{document}